\documentstyle[aps,preprint]{revtex}
\tightenlines
\begin{document}
\title{Simple relativistic model of a finite-size particle}
\author{Iwo Bialynicki-Birula}
\address{Centrum Fizyki Teoretycznej PAN
Lotnik\'ow 32/46, 02-668 Warsaw, Poland$\,^*$\\
and Institut f\"ur Theoretische Physik, Johann Wolfgang
Goethe-Universit\"at\\ Robert-Mayer-Strasse 8-10, Frankfurt am
Main, Germany}

\maketitle
\begin{abstract}
Soluble model of a relativistic particle describing a bag of matter
with fixed radius held together in perfect balance by a
self-consistent combination of three forces generated by
electromagnetic and massive scalar and vector fields is presented.
For realistic values of parameters the bag radius becomes that of
a proton.
\end{abstract}
\vspace{1cm}

$^*$ Permanent address. E-mail address: birula@planif61.bitnet
\vspace{2cm}

\newpage
{\bf 1. Introduction}
\vspace{1cm}

``I never satisfy myself until I can make a mechanical model of a
thing''. These words of Lord Kelvin \cite{kel} describe best my
motivation to search for a simple relativistic mechanical model of
an extended charged particle. Since over the years the concept of
a mechanical model has undergone a substantial evolution, I do not
hesitate to include also fields as building blocks for "mechanical"
models.

In this Letter I present a very simple model of a relativistic
charged object (one may think of it as a proton or a nucleus) that
is solvable in terms of elementary functions. The model consists of
a swarm of particles endowed with three types of charges $e$,
$g_S$, and $g_V$ interacting in a self-consistent manner with three
relativistic fields. The constituent particles are occupying a
bounded region in space (a bag) and are described by a scalar
phase-space distribution function $f({\bf r},{\bf p},t)$. The
solution of the field equations in this model can also be given a
hydrodynamic interpretation in which case the bag becomes a droplet
of pressureless fluid interacting with the three fields. An
unexpected result of this investigation is the appearance of a
quantization condition from which the radius of the bag (or the
droplet) is determined.

Almost a century ago Poincar\'e \cite{poi} argued that to
counterbalance electrostatic repulsion inside a charged particle
one must introduce cohesive forces --- Poincar\'e stresses. In
order to comply with the requirements of relativity theory, these
stresses must possess a dynamics of their own, they must in Pauli's
words ``depend on physical quantities which are causally determined
by differential  equations'' \cite{pau}. Nowadays we know that a
possible candidate for a ``causally determined physical quantity''
that will hold a charged particle together is a scalar field. For
unlike the electrostatic forces generated by vector fields, the
forces between like charges generated by scalar fields are
attractive. To prevent collapse one needs also a short range
repulsion and that in turn can be supplied by a massive vector
field. That explains the choice of the main ingredients of my
model. They are, of course, being used quite often in particle
physics and in nuclear physics \cite{sw}.

Several classical models of a charged particle have been invented
in the past (see, for example, \cite{dir,sch,bb} and also the
reviews of the classical electron theory \cite{roh,mil,pea,ady}),
but in all of them the Poincar\'e stresses were introduced ad hoc.
The relativistic model of a classical charged particle with a
finite, sharply defined radius presented here is the first, to my
knowledge, in which the Poincar\'e stresses incorporate some
realistic elements. It can be used in two ways. First, with its
help one may illustrate and clarify some old problems of
relativistic theories of extended objects, discussed already by
Abraham \cite{abr}, Poincar\'e \cite{poi}, Lorentz \cite{lor}, and
von Laue \cite{lau,lau1}, that also more recently continued to
cause controversy \cite{boy,roh1}. Second, with the proper choice
of parameters, the model may serve as a zeroth order approximation
in theories of nuclei and their high-energy collisions. For it is
clear today that if we are to use in realistic applications a
semi-classical model of a relativistic extended object it will not
be to describe an electron as Poincar\'e and others had tried in
vain. We may still try, however, to describe in this manner
nucleons and nuclei whose quantum features are less predominant
since their Compton wave lengths are much smaller than their
physical dimensions.
\vspace{1cm}

{\bf 2. Description of the model}
\vspace{1cm}

The starting point of my construction is the following set of
relativistic equations describing the motion of a relativistic dust
of particles interacting in a self-consistent manner with the three
fields ($c = 1$)
\begin{eqnarray}
\!\!\![(m-g_S\phi)(\partial_t + {\bf v}\cdot\nabla)
+ m{\bf F}\cdot\mbox{\boldmath $\partial_p$}\,]\,f({\bf r},{\bf
p},t) &=&
0,\label{eq1}\\
\partial_\mu F^{\mu\nu} &=& ej^\nu,\label{eq2}\\
(\raisebox{-.3ex}{$\Box$} + m^2_S)\phi &=& g_S\rho,\label{eq3}\\
\partial_\mu G^{\mu\nu} + m^2_V W^\nu &=& g_Vj^\nu.\label{eq4}
\end{eqnarray}
where $\nabla$ and \mbox{\boldmath $\partial_p$} denote the
derivatives with respect to ${\bf r}$ and ${\bf p}$, respectively.
The laboratory-frame three-velocity ${\bf v}$ and the three-force
${\bf F}$ are related to the spatial parts of the four-velocity
$u^\mu$ and the four-force $f^\mu$,

\begin{eqnarray}
u^\mu &=& p^\mu/m,\nonumber\\
f^\mu &=& e F^{\mu\nu}u_\nu - g_S(\partial^\mu - u^\mu
u^\nu\partial_\nu)\phi + g_V G^{\mu\nu}u_\nu,\label{force}
\end{eqnarray}
in the usual way, namely, $v^i =  u^i/u^0 = p^i/E_p, \; F^i =
f^i/u^0$. The antisymmetric field tensors $F_{\mu \nu}$ and $G_{\mu
\nu}$ are constructed in the standard manner from the corresponding
potential four-vectors $A_\mu$ and $W_\mu$. The four-current
$j^\mu$ and the scalar density $\rho$ are defined in terms of the
distribution function as follows

\begin{eqnarray}
j^\mu({\bf r},t) = \int \!\frac{d^3p}{E_p}p^\mu f({\bf r},{\bf
p},t),\\
\rho({\bf r},t) = \int \!\frac{d^3p}{E_p}mf({\bf r},{\bf p},t).
\end{eqnarray}

The set of coupled equations (\ref{eq1})--(\ref{eq4}) is a
generalization of Vlasov-Maxwell equations used in plasma physics.
Despite the nonrelativistic appearance of Eq.(\ref{eq1}), full
relativistic covariance of the theory based on these equations can
be proven in the same manner as we have done earlier \cite{bht} for
the pure Vlasov-Maxwell theory.

Equations (\ref{eq1})--(\ref{eq4}) may be viewed as classical if
one assigns the parameters $m_S$ and $m_V$ a dimension of inverse
length and uses them only as measures of the range of forces due to
scalar and vector fields. Planck's constant will make its
appearance at the end in Eqs.(\ref{par1}) and (\ref{par2}) when the
parameters of the model are expressed in terms of the masses of
mesons and their coupling constants.
\vspace{1cm}

{\bf 3. A static solution of the model}
\vspace{1cm}

A solution describing a bag of finite size is most easily obtained
in the static case, when all constituent particles are at rest,

\begin{eqnarray}
f({\bf r},{\bf p},t) = \delta({\bf p})\,\rho({\bf r}),\label{sdf}
\end{eqnarray}
and the field equations (\ref{eq2})--(\ref{eq4}) reduce to the
following simple set (in the static case there is just one scalar
density $\rho$ since $E_p=m$)

\begin{eqnarray}
- \Delta A_0 &=& e\rho,\label{st1}\\
(- \Delta + m^2_S)\phi &=& g_S\rho,\label{st2}\\
(- \Delta + m^2_V)W_0 &=& g_V\rho.\label{st3}
\end{eqnarray}
In order to satisfy also the equation (\ref{eq1}) for the
distribution function (\ref{sdf}), I assume the following
equilibrium condition

\begin{eqnarray}
\rho {\bf F} \equiv -\rho\nabla(eA_0 - g_S\phi + g_VW_0) = 0.
\label{equ}
\end{eqnarray}
This condition simply means that the net force acting on each
particle in the bag vanishes. Note that the equilibrium condition
is not imposed everywhere in space --- that would lead to a trivial
solution --- but only in those regions of space where matter is
present. That is why the bag's boundary must be well defined. The
equilibrium condition severely restricts possible solutions, but
fortunately it does leave room for some interesting ones. Since the
solutions of the equations (\ref{st1})--(\ref{equ}) can always be
scaled, I shall normalize them by imposing the following
normalization condition

\begin{eqnarray}
\int d^3 r\,\rho = 1.
\end{eqnarray}

A simple, spherically symmetric solution of
Eqs.(\ref{st1})--(\ref{st3}) is written below separately for the
inside and the outside of the bag.

Inside ($r\leq R$):

\begin{eqnarray}
\rho &=& f_+ - f_-,\\
eA_0 &=& e^2\Bigl(\frac{f_+}{k^2_+} -\frac{f_-}{k^2_-}\Bigr) -
V_0,\\
g_S\phi &=& g^2_S\Bigl(\frac{f_+}{k^2_+ + m^2_S} -\frac{f_-}{k^2_-
+ m^2_S}\Bigr),\\
g_V W_0 &=& g^2_V\Bigl(\frac{f_+}{k^2_+ + m^2_V} -\frac{f_-}{k^2_-
+ m^2_V}\Bigr),
\end{eqnarray}

Outside ($r>R$):

\begin{eqnarray}
\rho &=& 0,\\
e A_0 &=& \frac{e^2}{4\pi r},\\
g_S\phi &=& \frac{b_S}{4\pi r} e^{-m_S(r-R)},\\
g_VW_0 &=& \frac{b_V}{4\pi r} e^{-m_V(r-R)},
\end{eqnarray}
where $b_S$, $b_V$, and $V_0$ are constants and $f_\pm$ are the
following S-wave solutions of the Helmholtz equations,

\begin{eqnarray}
f_\pm = \frac{d_\pm}{4\pi}\,\frac{\sin(k_\pm r)}{r}.
\end{eqnarray}
The wave vectors $k_\pm$ are determined from a biquadratic equation
obtained from the equilibrium condition (\ref{equ}),

\begin{eqnarray}
k^2_\pm = \frac{B \pm \sqrt{D}} {2},
\end{eqnarray}
where

\begin{eqnarray}
Q^2 = e^2 - g^2_S + g^2_V,
\end{eqnarray}

\begin{eqnarray}
D = B^2 -4e^2 Q^2 m^2_S m^2_V,
\end{eqnarray}
and

\begin{eqnarray}
B = (g^2_S-e^2)m^2_V - (g^2_V + e^2)m^2_S.
\end{eqnarray}
The remaining parameters $b_S, b_V$, $d_\pm$, the depth of the
potential well $V_0$, and the radius of the bag $R$ are determined
from six continuity conditions at the bag's boundary. A combination
of these conditions gives the following quantization condition for
$R$, very similar to those arising in wave mechanics,

\begin{eqnarray}
T(m_S) = T(m_V),\label{qc}
\end{eqnarray}
where

\begin{eqnarray}
T(m_X) = \frac{k^2_+ + m^2_X}{k^2_- + m^2_X}\;\frac{k_- +
m_X\tan(k_-R)} {k_+ +m_X\tan(k_+ R)}.\label{dfd}
\end{eqnarray}
Eq.(\ref{qc}) has infinitely many solutions for $R$, but only the
lowest one is physically acceptable because all higher ones do not
lead to a positive density $\rho$. Once the radius of the bag is
determined, the remaining parameters can be calculated from
explicit formulas. In particular,

\begin{eqnarray}
V_0 = \frac{e^2}{4\pi}\;\frac{Tk_+ - k^2_+/k_-}{Tt_+ - k^2_+
t_-/k^2_-},
\end{eqnarray}
where $t_\pm = \tan(k_\pm R) - k_\pm R\;$ and $T$ is the common
value of $T(m_S)$ and $T(m_V)$.

The solutions of field equations obtained in this way have several
general features that are worth noting. The bag has a sharp edge
--- the value of the density at the surface of the bag is always
finite,

\begin{eqnarray}
\rho(R) = \frac{e^2 m_S m_V}{4\pi Q^2 R}.
\end{eqnarray}
It is seen from this formula that for a solution to exist the
effective repulsion must be stronger than attraction ($Q^2>0$).
\vspace{1cm}

{\bf 4. Hydrodynamic description}
\vspace{1cm}

My bag model may easily be converted into a droplet model by
replacing the description in terms of the distribution function by
a hydrodynamic description in terms of a density scalar field
$\rho$ and a four-velocity field $u^\mu$ that characterize the
state of the fluid. The starting point of the hydrodynamic
description is the following set of relativistic equations

\begin{eqnarray}
\partial_\mu(\rho u^\mu) &=& 0,\label{heq1}\\
(m - g_S\phi)\rho u^\nu\partial_\nu u^\mu &=& e\rho F^{\mu\nu}u_\nu
- g_S\rho(\partial^\mu\phi - u^\mu u^\nu \partial_\nu\phi)+ g_V\rho
G^{\mu\nu}u_\nu,\label{heq2}\\
\partial_\mu F^{\mu\nu} &=& e\rho u^\nu,\label{heq3}\\
(\raisebox{-.3ex}{$\Box$} + m_S^2)\phi &=& g_S\rho,\label{heq4}\\
\partial_\mu G^{\mu\nu} + m_V^2 W^\nu &=& g_V\rho
u^\nu.\label{heq5}
\end{eqnarray}
These equations describe matter modelled by a pressureless fluid of
density $\rho$ moving with four-velocity $u^\mu$. The fluid is
endowed with the three types of charges $e, g_S$, and $g_V$ and is
interacting in a self-consistent manner with the three relativistic
fields. For a fluid without pressure the static solution remains
the same as in the bag model. One may also add the pressure term
and an equation of state and seek numerical solutions of the same
general nature --- with a sharply defined radius.

The system described by the equations (\ref{heq1})--(\ref{heq5}) is
conservative (and that is also true for the system described by
generalized Vlasov-Maxwell equations (\ref{eq1})--(\ref{eq4}))
since the equations of motion guarantee that the total
energy-momentum tensor $T^{\mu\nu}$ of the system,

\begin{eqnarray}
T^{\mu\nu} &=& (m - g_S\phi)\rho u^\mu u^\nu + F^{\mu\lambda}
F_\lambda^{\ \nu} + {\textstyle {1\over 4}}
g^{\mu\nu}F_{\lambda\rho} F^{\lambda\rho} +
\partial^\mu\phi\partial^\nu\phi - g^{\mu\nu}{\textstyle {1\over
2}} (\partial_\lambda\phi\partial^\lambda\phi -
m_S^2\phi^2)\nonumber\\
&+& G^{\mu\lambda} G_\lambda^{\ \nu} + m_V^2 W^\mu W^\nu +
g^{\mu\nu}({\textstyle {1\over 4}}G_{\lambda\rho} G^{\lambda\rho}
- {\textstyle {1\over 2}}m_V^2 W_\lambda W^\lambda),\label{emt}
\end{eqnarray}
is conserved, $\;\partial_\mu T^{\mu\nu} = 0.$

The droplet (or the bag) described here is energetically stable.
The combined energy of the three fields, owing to the equilibrium
condition, reduces to half of the interaction energy with the
scalar field (a version of the virial theorem) and that leads to
the following value of the total energy of the system obtained from
(\ref{emt}) by integrating the energy density

\begin{eqnarray}
E_{tot} &=& \int\!d^3r(m - g_S\phi)\rho +
{\textstyle {1\over 2}}\int\!d^3 r[(\nabla A_0)^2 + (\nabla\phi)^2
+ m^2_S\phi^2 + (\nabla W_0)^2 + m^2_V W_0^2]\nonumber\\
&=& \int\!d^3r(m - g_S\phi)\rho + {\textstyle {1\over 2}}\int\!d^3
r[-A_0\Delta A_0 - \phi\Delta\phi + m^2_S\phi^2 - W_0\Delta W_0 +
m^2_V W_0^2]\nonumber\\
&=& m + {\textstyle {1\over 2}}\int\!d^3 r[eA_0 \rho - g_S\phi \rho
+  g_V W_0\rho] = m - V_0/2.
\end{eqnarray}
This formula conveys an important message concerning equilibrium
configurations in local field theories. In order to calculate the
total energy of the matter-field configuration in the whole space
it is enough to know the field inside the bag.
Since we know the relativistic transformation properties of the
fields describing the model, a static solution may easily be
boosted to give a model of a moving particle. The infamous factor
of 4/3 discovered by J.J.Thompson \cite{thom} that was plaguing all
naive electron models does not appear here. It can be checked by an
explicit calculation or inferred from a generalized virial theorem
(cf., for example, \cite{bb}) that my solution satisfies the von
Laue's condition \cite{lau1}

\begin{eqnarray}
\int\!d^3r\, T^{ii}({\bf r}) = 0,
\end{eqnarray}
that guarantees proper transformation properties of the
energy-momentum vector of the particle.
\vspace{1cm}

{\bf 5. A tentative connection with reality}
\vspace{1cm}

Now I come to the problem of assigning definite values to the
parameters of the model in order to see if the bag can be made to
resemble a nucleon. I have not done any elaborate parameter
fitting, since that will definitely require the introduction of
several scalar and vector mesons as is done in modern mean-field
theories of nuclear structure \cite{sw}. I have just taken as my
input the same values of the masses of the $\sigma$-meson and the
$\omega$-meson and the values of the coupling constants that are
used in the simplest version (QHD-I) of the mean field theory of
nuclear matter (Ref.\ \cite{sw},p.\ 125),

\begin{eqnarray}
m_S &=  550\;{\rm MeV}/\hbar c,\;\; m_V &= 783\;{\rm MeV}/\hbar c,
\label{par1}\\
g^2_S &= 91.64\;\hbar c,\;\;\;\;\;\;\;\;\;\;g^2_V &= 136.2\;\hbar
c.\label{par2} \end{eqnarray}
The quantization condition (\ref{qc}) gives then the following
values for the radius of the bag and the depth of the binding
potential

\begin{eqnarray}
R = 1.05\;{\rm fm},\;\;\;V_0 = 31.42\;{\rm MeV}.\label{par3}
\end{eqnarray}
The values of the proton mean radius and the binding energy
calculated for these values are
\begin{eqnarray}
r_p = 0.714\;{\rm fm},\;\;\;E_B = 15.71\;{\rm MeV},
\end{eqnarray}
and they are to be compared with the experimentally measured radius
0.862 fm and the bulk binding energy per nucleon in nuclear matter
15.75 MeV. I am very far from advocating the use of this model in
its present rudimentary form as a realistic model of nuclear
structure and I am presenting these numbers only to show that they
are not completely out of touch with reality.

To illustrate my results I show in Fig.\ \ref{fig1} separately the
curves for the three potentials, in Fig.\ \ref{fig2} the combined
effective potential well, and in Fig.\ \ref{fig3} the distribution
of matter in the bag. All these graphs correspond to the choice of
parameters given by the formulas (\ref{par1}) and (\ref{par2}).

The potential well shown Fig.\ \ref{fig2} may be used as an initial
step in constructing a relativistic version of the nuclear shell
model in which the Dirac equation would be used to describe matter.
More ambitiously, one may use the understanding reached with my
simple model (in particular, the quantization conditions for the
radius) to solve more elaborate models in which matter is also
described at the field theoretic level.
\vspace{1cm}

{\bf Acknowledgment}
\vspace{1cm}

I would like to acknowledge the hospitality extended to me by
Walter Greiner at the University of Frankfurt.

\begin{figure}
\caption{Separate graphs of the three self-consistent potentials
plotted for the values of the parameters given by
Eq.(\protect\ref{par1}) and Eq.(\protect\ref{par2}). Two repulsive
potentials, long range electrostatic (owing to the smallness of the
fine structure constant barely different from zero on this scale)
and short range mesonic, are due to the vector fields. The
attractive potential is due to the scalar field.}
\label{fig1}
\end{figure}

\begin{figure}
\caption{Total effective potential due to the combined effect of
the three fields producing a potential well that keeps the
constituents inside the bag.}
\label{fig2}
\end{figure}

\begin{figure}
\caption{The density of matter $\rho$ for a spherically symmetric
bag plotted as a function of $r$. Note a discontinuity of the
density at the bag's boundary.}
\label{fig3}
\end{figure}

\end{document}